\documentclass[12pt]{article}
\usepackage{latexsym} \usepackage{epsf}
\usepackage{epsfig}
\usepackage{a4}
\usepackage{amsfonts}
\usepackage[colorlinks]{hyperref}
\usepackage{color}
\usepackage{hyperref}
\newcommand{\ba}{\begin{array}}
\newcommand{\ea}{\end{array}}
\newcommand{\be}{\begin{equation}}
\newcommand{\ee}{\end{equation}}
\newcommand{\nn}{\nonumber}
\newcommand{\bea}{\begin{eqnarray}}
\newcommand{\ena}{\end{eqnarray}}
\newcommand{\beas}{\begin{eqnarray*}}
\newcommand{\enas}{\end{eqnarray*}}
\newcommand{\bq}{\begin{quotation}}
\newcommand{\eq}{\end{quotation}}

\renewcommand{\theequation}{\thesection.\arabic{equation}}

\begin{document}

\begin{center}
{\Large{ Integrability in $[d+1]$ dimensions: combined local
equations and commutativity of the transfer matrices}}
\end{center}

\begin{center}{\bf
Shahane A. Khachatryan \footnote{e-mail:{\sl shah@mail.yerphi.am}
 }}
 \\ A.~I.~Alikhanyan National Science Laboratory (YerPhI), 
 \\ Alikhanian Br. str. 2, Yerevan 36, Armenia
\end{center}

\begin{abstract}

 We propose new
  inhomogeneous local integrability equations - combined equations,
 for
 statistical vertex  models of general dimensions in the framework of the Algebraic Bethe Ansatz (ABA).  For the
 low dimensional cases  the efficiency  of the
step by step consideration of the transfer matrices'
commutation is demonstrated.
 We construct some simple 3D
solutions with the three-state $R$-matrices of certain 20-vertex structure; the connection with the quantum
three-qubit gates is discussed. New, restricted versions of  3D local integrability equations with four-state $R$-matrices are defined, too. Then  we construct  a new 3D analogue
 of the two-dimensional star-triangle equations.
\end{abstract}


{\small\tableofcontents}
\newpage
\section{Introduction}

The Algebraic Bethe Ansatz (ABA)  is a powerful technique  developed for  investigating integrable models in
$[1+1]$D quantum theory or $2$D classical statistical mechanics, \cite{Y}-\cite{GSA}. The models discussed  by means of ABA are specified using local weights or $R$-matrices that satisfy spectral-dependent Yang-Baxter  Equations  (YBE), the key ABA relations.  The YBEs  ensure the commutativity of the transfer
matrices constructed by  $R$-matrices with different spectral
parameters.
The ABA offers an algebraic way  for constructing the ground state wave function, energy
spectra and self-functions of the Hamilonian (\cite{FT}, \cite{ZZ1}, \cite{KBI}). There are several
generalizations of  YBE proposed for 3D case. The first and  most famous  equations
are Zamolodchikov's Tetrahedron Equations (ZTE)   imposed on the  scattering  operators of a $[1+1]$-dimensional  object - a string \cite{ZZ}.  In papers  \cite{Baxt83,BBt,Kor,Hie,KMS} the  interaction-round-cube (IRC) and  vertex versions of ZTE were formulated with
solutions equivalent to Zamolodchikov's trigonometric two-state solution
 or being its generalizations for the $N$-state case.
 For the modified ZTE,  elliptic generalizations for two-layer transfer matrices  were constructed, see  \cite{MS}-\cite{Hu1}. The paper \cite{ASHS} suggested simplified or semi-tetrahedron equations. In a
recent work \cite{KhFKS},  another set of local equations  was proposed for three-dimensional lattice models - the cube
equations, which can also provide commutative  transfer
matrices, with a solution 
 corresponding to the
quantum integrable 2D model of Kitaev \cite{Kitaev}. However, it is not an easy task to obtain solutions that could produce local and Hermitian quantum  two-dimensional  Hamiltonian operators.  For all cases, the main problem is that
there are a huge number of  equations and undefined functions of various
spectral variables, and  only a limited number of solutions are obtained that
satisfy the conditions of three-dimensional local integrability. ZTEs are the most
analyzed and  productive equations \cite{ZZ,Zt}.  A big step forward in the study of   integrability in 2D appeared  to be the exploration of the symmetry properties of
 models (conformal symmetry, quantum group), \cite{QGR}-\cite{isa}, \cite{shk}-\cite{SHhk}.
  For the three-dimensional  case also, there are some investigations  of models with
  quantum group symmetry (for example \cite{BVVSS}). The Ansatz technique can be extended to higher dimensions, too. Taking into account the established connection between $[d+1]$ space-time quantum theories and  $\mathbb{D}\equiv(d+1)$ dimensional classical statistical physics \cite{Baxt}, the higher-dimensional statical weights $R_{\mathbb{D}}$, being the solutions to the homogeneous generalizations of tetrahedral equations, known in the literature as $\mathbb{D}$-simplex equations, can be used  for constructing quantum integrable models' Hamiltonian operators on $d$ dimensional regular lattices \cite{ZZ,Zt,BBt1}.

 In this manuscript we analyze integrability questions in the context of ABA from two different points of view, generally  described in the Subsection 2.1 and Subsection 2.2. We propose new local integrability equations (non-homogeneous and simplified) - Subection 2.1, Sections 4,5,6, alternative to the existing ones, and together with it, we discuss the possibility to explore
 the transfer matrices' commutativity, avoiding direct
consideration of the local YBE-type equations - Subection 2.2, Section 3. 
  In Section 2, we define and discuss
  non-homogeneous
 local compatibility equations for the general case of $[d+1]$ .
   In  Subsection 2.1, we suggest a  realization of the YBE-type local equations for higher dimensional vertex models -  combined  equations or restricted $\mathbb{D}$-simplex equations, with a possible variety of the intertwiner matrices.  In contrast to  homogeneous  $\mathbb{D}$-simplex equations, where all vertices and, hence,
    matrices are considered equal, in the discussed  inhomogeneous equations the intertwiner matrices can be of smaller dimension than the quantum $R$-matrices. In the simplest case, all the intetwiner matrices can be taken as $R$-matrices of minimal dimension  two (i.e., acting on the tensor product of two states).
     As it is noted already, such simplified equations were first proposed for 3D in \cite{ASHS}.
  In the works \cite{AKS,ASAR} the plain version of three-state $R_3$-matrices was discussed,
  where  irreducible 3D matrices were used in the
  construction of  integrable one-dimensional  quantum chain spin models.

   In  Subsection 2.2 we demonstrate how one can construct integrable models
via deducing the matrix elements of
   $R$-matrices just from the integrability conditions followed from the commutations of the
   transfer matrices with a small number $n$ of  lattice sites, without fixing local integrability relations.
We  see that step-by-step consideration of such equations can completely solve the problem by analyzing only a limited number of $n$-s. The value of $n$ is conditioned
  by different factors, for example,  the number of state characteristics (e.g., dimensions), the number of  non-zero elements of  $R$-matrices.
   An apparent advantage and hence the  purpose of the considered point of view is that we do not fix the kind of the local equations, which is essential for  $\mathbb{D}>2$. Then such equations are  homogeneous  in respect to the matrix-element functions  depending on the spectral
  variables $\{u\}$ and $\{v\}$ (these can be compound variables in general, see e.g.  \cite{SHY}). Meanwhile,  the Yang-Baxter equations in 2D, and their analogs for general $[d+1]$ space-time \cite{BBS1}
    contain also the matrix element-functions of the intertwiner matrices  depending from the variables $\{u,\;v\}$ (and more).  As an example, by the described scheme  we easily reveal the most general non-homogeneous
 eight-vertex model's
    $R$-matrices (see the Appendix), which exactly coincide with the previously obtained results in
    \cite{Baxt,KhS,SHY}.
     In  Section 3 we  consider some
      simple  R-matrices for the 3D $20$-vertex models,
       represented by the projection
      operators. We verify  the existence of the solutions by the
       consideration of the vertex-like
       ZTE and their simplified versions.
        Some simple new 3D statistical integrable models
         are presented in Section 3. The connection of  the investigated 3D matrices with the three-qubit quantum gates \cite{LK}-\cite{EDZ}
         is discussed in the following section. In the next two
       sections we construct
        the combined versions  of the  IRC o 12-hedron equations \cite{Baxt83,BBt}
          (an analogue of 2D face type YBE, Section 4) and
            the combined cube equations \cite{KhFKS}
          (Section 5).  Here the cube  equations are modified and presented
in the so-called “semi-check” formulation, with slight adaptations of the
indexing and with the change of the “time” direction.

 As another version of the local integrability   equations in 3D, we  propose in Section 6  a new generalization
 of the star-triangle equations, which can be referred as  “connected” star-triangle relations.
  The presented simple solutions are inherited from the two-dimensional solutions  in quite a natural way.


\setcounter{equation}{0}
\section{The transfer matrices' commutation conditions}

In the context of the Algebraic Bethe Ansatz (\cite{Y}-\cite{KBI})
 integrable $[1+1]$-lattice quantum models can be  described by
means of the  monodromy  matrix $T(u)$
constructed by the product of local $R_{ai}(u)$-matrices, depending on spectral parameters, $R_{ai}:V_a\otimes V_i$,
\bea T(u)_{\{a\};i_1...i_N}=R_{a\;i_1}(u)R_{a\;i_2}(u)...R_{a\;i_N}(u),\qquad \tau(u)=tr_{\{a\}}[T(u)_{\{a\};\{i\}}]. \label{Tua}\ena
 The  operator $T(u)$ acts on the tensor product of the auxiliary space  $V_a$ and  the quantum spaces $V_{i_1}$ defined on  a
  chain with $N$
 sites - $V_a\otimes V_{i_1}\otimes V_{i_2}....\otimes V_{i_N}$. The transfer matrix $\tau(u)$ in some sense plays a role of discrete evolution operator in $(1+1)$ space-time.  The key integrability condition is the commutativity of the
  transfer matrices with different spectral parameters
\bea[\tau(u),\tau(v)]=0.\label{ttu}\ena
   This gives an opportunity to construct the full spectra (sufficient for integrability of the problem)  of the
 conserved charges, generated by the transfer matrix.  These mutually commutative operators emerge in the
  spectral parameter expansion of the transfer
matrix.  Particularly the Hamiltonian operator is expressed by the logarithmic
 derivative $H\simeq\frac{\partial_u\tau(u)}{\tau(u)}|_{u=u_0}$. For ensuring the locality of the models, here the point $u_0$ is chosen so, that $R(u_0)\equiv P$ is a permutation operator. The spectral parameter $u$ plays a role of
 the time evolution parameter, and it is attached to the auxiliary space $V_a$. Also one can attach to each quantum space $V_i$ a local spectral parameter $w$, $R_{a\;i}(u,w)$.
  The main part of the discussed $R$-matrices behaves a difference property: $R(u,w)=R(u-w)$, in terms of the additive parameters $u,\;w$.

 The higher dimensional   generalization of ABA can be realised on the homogeneous $N\times N\times ...\times N$ regular lattices with toric configurations, i.e. for the quantum  models defined on the lattice with $\mathbf{N}=N^d$ sites, and  having periodic boundary conditions \cite{Zt,BBS1}. For the vertex-type models, for which the $R$-matrices are attached to the vertexes, and the states - to the links, the generalization is obvious.  The quantum states for the vertex models are disposed on the links along the time axe (e.g., the links indexed by $i_z,\; j_z$ on the Fig. 1), and the auxiliary states are attached to the links directed along the space axes (e.g., the links indexed by $i_{x,y},\; j_{x,y}$ in Fig. 1). Now at each vertex, described by $d$-dimensional vector $\vec{k}$, there are defined auxiliary states of number $d$:  $V_{\mathbf{a}}=\bigotimes_{j}^d V_{a^j}$,  and the auxiliary index $\mathbf{a}$ in
 $R_{\mathbf{a}_{\vec{k}}i_{\vec{k}}}$ must be understood as a compound index $\mathbf{a}=a^1,...,a^{d}$, and
 correspondingly, the spectral parameter $u$ now can be consider as $d$-dimensional parameter $\bf{u}=\{u^1,...,u^d\}$.
 The  vector-index ${\vec{k}}$  is described by means of the coordinates $\{k^1,...,k^d\}$.  Each vertex has  $[d+1]$
 incoming and $[d+1]$ outgoing links, corresponding to the down and upper indexes of the $R$-matrices:
 ${[R_{\mathbf{a}i_{\vec{k}}}]_{{a^1_{\vec{k}}...a^d_{\vec{k}}i_{\vec{k}}}}^{b^1_{\vec{k}}...b^d_{\vec{k}}j_{\vec{k}}}}$.
 Then the transfer
 matrix must be defined over  the product of the all $\mathbf{N}$   quantum spaces. The trace  in (\ref{Tua}) is taken
 over all the  auxiliary spaces of the number
  \be \mathbf{N}_e=d\times N^{d-1},\ee
  which is the number of the boundary states. So, in the monodromy matrix 
   now  $\{a\}\to \{\mathbf{a}_{\vec{{\bf{k}}}_e}\}=\{a^1_{\vec{k}_e},...,a^{\mathbf{N}_e}_{\vec{k}_e}\}$, where the boundary
   states are described by the boundary vector indexes $\vec{k}_e$, for which at least one of the coordinates equals to
   $1$ (or $N$); $\{k_e^1,...k_e^n,....,k_e^d\}=\{r^1,...,1,...,r^d\}$, $r^k\in{1,...,N}$. Below for simplicity we shall omit in $\vec{k}_e$ the index
   $e$, denoting the boundary.

{\unitlength=0.25cm
\begin{figure}[t]
 \begin{picture}(100,20)
 \put(10,0){\line(1,0){10}}\put(8.9,-1){$a$}
\put(20,0){\line(0,1){10}}\put(20.5,-1){$f$}
 \put(10,10){\line(1,0){10}}\put(8.9,9.2){$e$}
\put(10,0){\line(0,1){10}}\put(20.5,9.2){$d$}
 \put(10,0){\line(2,1){8}}
\put(10,0){\line(1,0){10}} \put(20,0){\line(2,1){8}}
\put(18,4){\line(1,0){10}}\put(18,4){\line(0,1){10}}
\put(28,4){\line(0,1){10}}\put(18,14){\line(1,0){10}}
\put(10,10){\line(2,1){8}}\put(20,10){\line(2,1){8}}
\put(16.8,4){$g$}\put(28.5,4){$b$}\put(16.8,14.2){$c$}\put(28.5,14.2){$h$}

\put(40,12){$i_x=-a+c,$}\put(40,8){$i_y=e-f,$}\put(40,4){$i_z=a-b,$}
\put(51,12){$j_x=-f+h,$}\put(51,8){$j_y=-b+c,$}\put(51,4){$j_z=e-h.$}
\put(38,0){$W(a|efg|bcd|h) \quad \rightarrow \quad
R_{i_x i_y i_z}^{j_x j_y j_z}.$}
\put(10,0){\line(1,0){10}}
\put(15,5){\vector(2,1){8}}\put(14.2,3.9){$i_y$}\put(22.9,10){$j_y$}\multiput(14.7,4.7)(8,4){2}{$\bullet$}

\put(14,7){\vector(1,0){10}}\put(12.5,7){$i_x$}\put(24.4,7){$j_x$}\multiput(13.7,6.6)(10,0){2}{$\bullet$}
\put(19,2){\vector(0,1){10}}\put(18,1){$i_z$}\put(19.7,12){$j_z$}\multiput(18.7,1.7)(0,10){2}{$\bullet$}
\end{picture}
\caption{Vertex and IRC weight functions (an example of
correspondence). In the cube equations this weight figures as $R_{afgb}^{edch}$-matrix.}\label{Figv-f}
\end{figure}}

\paragraph{Note,} that in the described formulation, the indexing of $3$D vertex matrix appears
in the following way: $R_{a^1_{\vec{k}}a^2_{\vec{k}}i_{\vec{k}}}^{b^1_{\vec{k}}b^2_{\vec{k}}j_{\vec{k}}}$, $\vec{k}=
\{k_x,k_y\}$, instead of the usual formula $R_{i_x i_y i_z}^{j_x j_y j_z}$ brought in the Fig. 1.

  \subsection{Non-homogeneous local integrability equations}

 The existence of the intertwiner matrix $\bar{R}_{\mathbf{a}\mathbf{b}}(u,v)$, which is an  invertible matrix in general case, satisfying to the following relations,
\bea
\bar{R}_{\{\mathbf{ab}\}}({\mathbf{u},\mathbf{v}})T_{\{\mathbf{a}\}\{\mathbf{i}\}}(\mathbf{u})T_{\{\mathbf{b}\}\{\mathbf{i}\}}(\mathbf{v})=
T_{\{\mathbf{b}\}\{\mathbf{i}\}}(\mathbf{v})T_{\{\mathbf{a}\}\{\mathbf{i}\}}(\mathbf{u})\bar{R}_{\{\mathbf{ab}\}}({\mathbf{u},\mathbf{v}}),
\label{tybe}\ena
will ensure the  commutativity of the  transfer matrices (\ref{ttu}). These can be considered as global integrability conditions. They automatically take place, if  $R(u)$-matrices satisfy  the following local relations -  spectral parameter dependent YBE-type equations
\bea&
\bar{R}_{\mathbf{a}_{\vec{k}}\bf{b}_{\vec{k}}}({\mathbf{u},\mathbf{v}})
R_{\mathbf{a}_{\vec{k}}{i}_{\vec{k}}}({\mathbf{u}})R_{\mathbf{b}_{\vec{k}}{i}_{\vec{k}}}({\mathbf{v}})=\!
R_{\mathbf{b}_{\vec{k}}{i}_{\vec{k}}}({\mathbf{v}})R_{\mathbf{a}_{\vec{k}}{i}_{\vec{k}}}({\mathbf{u}})
\bar{R}_{\mathbf{a}_{\vec{k}}\mathbf{b}_{\vec{k}}}({\mathbf{u},\mathbf{v}}).&\label{ybeu}\ena
The forms of the equations (\ref{tybe}, \ref{ybeu}) are usual for $(1+1)$-D case, however in principle they are
valid also for higher $[d+1]$ dimensions, %
 indicating in the formulas by  $d$-dimensional vector indexes $\vec{k}$. For three dimensional case the detailed
 proof is brought in \cite{ASHS}. In such cases the matrix $\bar{R}_{\mathbf{a\;b}_{{\vec{k}}}}(\mathbf{u},\mathbf{v})$
 must be factorized into
 $d$  interwiner matrices (\ref{ybe1}), and hence it will lead to the factorization of the global intertwiner matrix in
 (\ref{tybe})
 \bea &
\bar{R}_{\mathbf{a}_{\vec{k}}\mathbf{b}_{\vec{k}}}({\mathbf{u},\!\mathbf{v}})
=\prod_{p=1}^d\bar{R}_{{a}^p_{\vec{k}}{b}^p_{\vec{k}}}(u^p,\!v^p),\quad
\bar{R}_{\{\mathbf{ab}\}}({\mathbf{u},\mathbf{v}})=
\prod_{\{\vec{k},p\}}^{\mathbf{N}_e}
\bar{R}_{{a}^p_{\vec{k}}{b}^p_{\vec{k}}}(u^{p},\!v^{p}).&
\label{ybe1}\ena
For checking the commutativity one must place the unity operator
$I=\bar{R}[\bar{R}]^{-1}$ in the trace of the product of the
transfer matrices and repeatedly use the  local equations
(\ref{ybeu}),
\bea &tr_{\{\mathbf{a}\}}[\prod_{\{\vec{k}\}}^{\mathbf{N}}
R_{\mathbf{a}_{\vec{k}}i_{\vec{k}}}({\mathbf{u}}) 
tr_{\{\mathbf{b}\}} [\prod_{{\vec{k}}}^{\mathbf{N}}  R_{\mathbf{b}_{\vec{k}}{j}_{\vec{k}}}({\mathbf{v}})]
=tr_{\{\mathbf{a,b}\}}[\prod_{\{\vec{k}\}}^{\mathbf{N}}  \left(R_{\mathbf{a}_{\vec{k}}i_{\vec{k}}}({\mathbf{u}})
R_{\mathbf{b}_{\vec{k}}{i}_{\vec{k}}}(v)\right)]=\label{rtt}&\nn\\&
tr_{\{\mathbf{a,b}\}}[[\bar{R}]^{-1}_{\{\mathbf{ab}\}}({\bf{u},\bf{v}})
\bar{R}_{\{\mathbf{ab}\}}({\bf{u},\bf{v}})\prod_{\vec{k}}^{\mathbf{N}}
\left(R_{{\mathbf{a}}{i}_{\vec{k}}}({\mathbf{u}})
R_{\mathbf{b}{j}_{\vec{k}}}({\mathbf{v}})\right)]=&
\\\nn&tr_{\{\mathbf{a,b}\}}[[\bar{R}]^{-1}_{\{\mathbf{ab}\}}({\mathbf{u},\mathbf{v}})
\prod_{\{\vec{k}\}}^{\mathbf{N}_e}
R_{\mathbf{a}_k  i_{k}}({\mathbf{v}})
R_{\mathbf{b}_k i_{k}}({\mathbf{u}})\bar{R}_{\mathbf{a}_k\mathbf{b}_k}({\mathbf{u},\mathbf{v}}){\prod'}_{\{\vec{k}'\}}
\left(R_{\mathbf{a}_{\vec{k}'}{i}_{\vec{k}'}}({\mathbf{u}})
R_{\mathbf{b}_{\vec{k}'}{i}_{\vec{k}'}}({\mathbf{v}})\right)]=...=&\\\nn&
tr_{\{\mathbf{a},\mathbf{b}\}}[[\bar{R}]^{-1}_{\{\mathbf{ab}\}}(\mathbf{u},\mathbf{v})\prod_{\{\vec{k}\}}^{\mathbf{N}}
\left(R_{\mathbf{a}{i}_{\vec{k}}}({\mathbf{v}})
R_{\mathbf{b}{i}_{\vec{k}}}({\mathbf{u}})\right)\bar{R}_{\{\mathbf{ab}\}}({\mathbf{u},\mathbf{v}})]=
tr_{\{\mathbf{a,b}\}}[\prod_{\{\vec{k}\}}^{\mathbf{N}}
R_{\mathbf{a}_{\vec{k}}{i}_{\vec{k}}}({\mathbf{v}}) R_{\mathbf{b}_{\vec{k}}{i}_{\vec{k}}}({\mathbf{u}})]& \ena
   The product  ${\prod}'$ is defined as ${\prod'}_{\{\vec{k}'\}}\prod_{\{\vec{k}\}}^{\mathbf{N}_e}=\prod_{\{\vec{k}\}}^{\mathbf{N}}$. Here we present this detailed derivation, as this is a general principle for all types of the local integrable relations and for general
 dimensions. We have used the relations (\ref{ybeu}) for all the edge indexes ${\{{\mathbf{a}}_{\vec{k}},\;{\mathbf{b}}_{\vec{k}}\}^{{\mathbf{N}}_e}}$ simultaneously. In principle the intertwiners of number $d$ can differ one from another by their structure, and one can emphasise this by an additional index $\bar{R}^{p}_{{a^p b^p}}(u^p,v^p)$.  The detailed local equations now read as
\bea
&\bar{R}^1_{{a^1 b^1}}(u^1,v^1)\cdot\cdot\cdot\bar{R}^d_{{a^d b^d}}(u^d,v^d)R_{a^1...a^d\;{i}}(u^1,...,u^d,w)R_{b^1 ... b^d\;{i}}(v^1,...,v^d,w)=&\nn\\
&R_{b^1 ... b^d\;{i}}(v^1,...,v^d,w)R_{a^1 ... a^d\;{i}}(u^1,...,u^d,w)\bar{R}^d_{{a^d b^d}}(u^d,v^d)\cdot\cdot\cdot\bar{R}^1_{{a^1 b^1}}(u^1,v^1).&\label{ybed}\ena
Here, and in the most equations below, we are omitting the notation $\vec{k}$ for the coordinates of the vertices.  Such YBE-type
local equations, where the intertwiner matrices are defined on the tensor product of two states $V_{a^p}\otimes V_{b^p}$,
 and the quantum matrices are defined on the product of $[d+1]$ states: $V_{a^1}\otimes V_{a^2}...\otimes V_{a^d}\otimes V_{i}$, we can refer as {\it{combined}}
 local integrability equations -  combination of the matrices of different dimensions (or the combination of 2D
 YBEs of number $d$). Another formulation for these equations can be the {\it{simplified}} $\mathbb{D}$-simplex equations, in the same way, as we have used for three dimensional case in
\cite{ASHS}, defining the simplified (or semi-) tetrahedron equations (STE).

In (\ref{ybed}), for completeness, in the set of the spectral parameters of the quantum matrices we involve an additional
 spectral parameter $w$, attached to the links of the quantum states $V_i$, $R_{\{a\},i}(\{u\},w)$.

 Also additional auxiliary states could be included in the definition of the   intertwiner operators.  For
   homogeneous integrability equations, which reproduce the $\mathbb{D}$-simplex geometry (as triangle and tetrahedral equations  at $\mathbb{D}=2,3$),
    the  quantum  and the intertwiner  matrices, $R_{\mathbf{a},{i}}$ and $\bar{R}_{a^p,b^p}(u,v)$,  must have the same number of the indexes and the same number of the spectral parameters. And to achieve  this, we must make the
 following
 generalization
  \bea& \bar{R}_{a^p,b^p}(u,v)\to
\bar{R}_{a^p,b^p,c^{p_1},...,c^{p_{d-1}}}(u,v,w^{p_1},...,w^{p_{d-1}}),\;p=1,...,d.&\ena
The addition of new auxiliary  indexes $\{c^i\}_{i=1}^{d'}$  and new spectral
 parameters $w^i,\quad i=1,...,d'$, corresponds to the introducing of the auxiliary spaces $V_{c_i}$, $(i=1,...,d')$, of number $d'=\frac{d(d-1)}{2}$, so that each two $\bar{R}$-matrices have only one coinciding index. And  $d'$ is the minimal number satisfying to this condition.  Such consideration  also  coincides  with the interpretation of the local integrability equations as the factorability  conditions
 of the scattering of spatial $[d-1]$-“strings” (the generalizations
 of the one-dimensional objects in the Zamolodhikov's tetrahedral equations).  At $d=2$ this corresponds to the vertex version of the system of ZTEs \cite{Y,ZZ,BBS1,Baxt,JM,Zt,BBt}. Of course, one can also define the equations with the intertwiner matrices of intermediate dimensions $\in(2,...,d+1)$, and such relations we can  call {“combined”} equations as well. In the case, when the intertwiner matrices with less dimensions are  themselves the solutions to YBE-type local equations, it can be used for obtaining the higher dimensional integrable models (the corresponding quantum $R$-matrices) %
 \cite{ASHS}. The used parametrization is quite general. Note, that in the literature for the significant part of obtained solutions there is  a natural parametrization of the $R$-matrices, formulated by means of the independent angles
  corresponding to the geometry of the local equations (two angles of the triangles - YBE, five angles of the tetrahedron - ZTE, e.g., \cite{ZZ,BBS1,Baxt,YMP,Zt,BBt,Kor,Hie}).

\paragraph{Directly diagonalizable  transfer matrices.} One separate point is the case of the directly diagonalizable  transfer matrices, i.e when it is possible by a direct operation to solve the
 eigen-problem of the transfer matrix: $\tau(u)=U_u \tau(u)^d U_u^{-1}$, so that $\tau(u)^d $ is a diagonal operator, and $U$ is an unitary operator. This happens, e.g., when the model
can be described by free particles (the  periodic quadratic action  is easy to diagonalize in the Fourier  transformation basis). Does this necessarily mean, that the transfer matrices with different spectral parameters must commute for such integrable models?   Symbolically, let us write the relation
\bea[\tau(u),\tau(v)]=[U_u \tau(u)^d U_u^{-1},U_v \tau(v)^d U_v^{-1}]\Rightarrow [\tau(u)^d,\tau(v)^d]=0, \quad
{\textrm{if}}\qquad U_u=U_{(u,v)}^d U_v.
\ena
 here the operator $U_{(u,v)}^d $ is a spectral-parameter dependent diagonal operator. This also can be read as
 \be U_u=U_u^d U_{0},\ee
where $U_{0}$ is a constant
unitary matrix,  $U_u^d=U_{(u,0)}^d$. It will be  a  sufficient condition for the commutation of the transfer matrices with different parameters.

 \subsection{The stepwise consideration of the transfer matrixes' commutation
equations
}

 If we do not apply the concrete local commutativity conditions (supposing existence of an intertwiner matrix), which are sufficient (but not necessary) conditions, ensuring the global commutativity, we can step by step apply the commutativity principle. For example,  at $d=1$, on the lattice with $N$ sites,  one can define incomplete transfer matrices $\tau_n(u)$ on the sub-chains of $n$ neighbouring sites with sub-monodromy matrices $T_{i;i_1...i_n}$, $n=1,2,3,...,N$, $\tau_N(u)\equiv\tau(u)$. It means
\bea
 [\tau_n(u),\tau_n(v)]=0,\quad \tau_n(u)=tr_a[T_{a;i_1...i_n}]=tr_a\prod_{k=1}^n R_{a i_k},\quad n=1,2,...,N. \label{ceq-n}
 \ena
 This also preserves the locality principle, as we start from the low $n$ cases ($n=1,2,...$).

  The equations (\ref{ceq-n}) contain the functions depending only on the variables $u$ or $v$.
 Although the $n$-th equations in (\ref{ceq-n}) contain  homogeneous equations of  n-th order in respect to the functions with the spectral parameters $u$ and $v$, however the complexity in some sense is compensated by the step by step consideration of the equations, starting from $n=2,3,...$. The case of $n=1$ is satisfied automatically. The equations at $n=2$ give  constraints on the matrix elements, solving of which  simplifies the next set of the equations at $n=3$, and so on. And one can find some interesting facts.  As example, for the most investigated eight-vertex model \cite{Baxt}
   it is enough to consider only the equations up to $n=4$ (see Appendix), which will fix all the constraints on  $R$-matrix elements necessary for the integrability, and hence will reproduce the known solutions. For some particular cases it is enough to analyse only the equations (\ref{ceq-n}) with $n=2,3$. And this also can be checked for the general $15$-vertex model \cite{SHY}.
    And all the solutions to (\ref{ceq-n}) in that cases admit existence of the corresponding intertwiner matrices as solutions to  non-homogeneous YBE.

 For $[d+1]$-dimensional cases, the indexes $i_k$, $a$ and $n,\; k$ in (\ref{ceq-n}) must be replaced
  with the corresponding $d$-dimensional numbers: $i_k\to \{\mathbf{i}_{\bf{k}}\}=\{i^1_{\bf{k}},...,i^d_{\bf{k}}\}$, in the same way - $a\to \mathbf{a}=\{a^1,a^2,...,a^d\}$, and the compound numbers $k\to {\bf{k}}$ and $n\to \mathbf{n}=\{n^1,...,n^d\}$, now describe the corresponding fragments of  $d$-dimensional lattices.

In the next section we examine the case for $d=2$ with a definite ($20$-vertex) structure of the matrices. For some cases we obtain the exact answers by direct stepwise consideration, and for other cases we use also combine equations as an
additional sufficient equipment for final proving the existence of the solutions (obtained for smallest sub-lattices) for all $n$-s.

\section{The case of 3D: 20-vertex models}
 \setcounter{equation}{0}

 Let us consider $N_x\times N_y\times N_z$ regular cubic lattice with the extension  of  2D
 six-vertex model. The $R$-matrix or the statistical weight of 3D statistical vertex model,
  can be defined for each vertex of the cubic lattice as follows
 \bea \label{R3}
R=R_{i_x i_y i_z}^{j_x j_y j_z} e_{i_x}^{j_x}\otimes
e_{i_y}^{j_y}\otimes e_{i_z}^{j_z}, \qquad i_x+i_y+i_z=j_x+j_y+j_z
\;[mod\;2]
 \ena
 Here by $e_{i_a}^{j_a}$, $a=x,y,z,$ we denote the  basic two-dimensional operators defined on the  spaces situated
 on the links (with the indexes $\vec{i},\; \vec{j}$),  connecting to the vertexes
 $\{x,y,z\}$  (Fig. \ref{Figv-f}). 
 The relations in (\ref{R3}), put  on the indexes $i_a,\; j_b$  of the non-zero elements of the matrices  (\ref{R3}),
  define the so-called $20$-vertex model. %
The three-state $R$-matrix now can be presented as
\bea
 R_{ijk}=\left(\ba{cc} \mathbf{R}_{0}^{0}&\mathbf{R}_{0}^{1}\\\mathbf{\mathbf{R}}_{1}^{0}&\mathbf{R}_{1}^{1}\ea\right),\qquad
\mathbf{R}_{a}^{a}=
 \left(\ba{cccc} R_{a00}^{a00}&0&0&0\\0&R_{a01}^{a01}&R_{a10}^{a01}&0\\0&R_{a01}^{a10}&R_{a10}^{a10}&0\\0&0&0&R_{a11}^{a11}\ea\right),\\
\mathbf{R}_{0}^{1}= \left(\ba{cccc}
0&0&0&0\\R_{001}^{100}&0&0&0\\R_{010}^{100}&0&0&0\\0&R_{011}^{101}&R_{011}^{110}&0\ea\right),\qquad
\mathbf{R}_{1}^{0}= \left(\ba{cccc}
0&R_{100}^{001}&R_{100}^{010}&0\\0&0&0&R_{101}^{011}\\0&0&0&R_{110}^{011}\\0&0&0&0\ea\right).\ena

The transfer matrix  reads as $\tau(\mathbf{u})=\sum
_{i=1,j=1}^{N_x,N_y} \prod_k R_{ijk}(\mathbf{u})$, where the product is taken over  2D square lattice on the flat surface, and taking the trace over the states with $i,j$ indexes of 2D $N_x\times N_y$ lattice implies that the periodic boundary conditions are imposed: $|V_{i+N_x,j+N_y}\rangle=|V_{i,j}\rangle$. The spectral parameter here consists of three components $\mathbf{u}=\{u_x,u_y,u_z\}$ (the “velocity” in the context of theory of
scattering), and as for two-dimensional case, the property of the additivity of the spectral parameter
allows to reduce their number by one:
$\{u_x,u_y,u_z\}\to \{u_x-u_z,u_y-u_z\}$. As it is analysed in \cite{SHY}, all the additional colored parameters of the $R$-matrix in general can be expressed in terms of the arbitrary functions arisen in the solutions of the equations.

Step by step consideration of the transfer matrices commutativity
here means that  the following relations take place:
\bea
[\tau_{n_x,n_y}(\mathbf{u}),\tau_{n_x,n_y}(\mathbf{v})]=0,\quad
\tau_{n_x,n_y}(\mathbf{u})\equiv tr_{i=1,j=1}^{n_x,n_y} \prod_k
R_{ijk}(\mathbf{u}),\quad n_x/n_y=1,...,N_x/N_y. \label{ttun}
\ena
At $n_x=1$ or
$n_y=1$, taking the trace over the variables attached at the one of the auxiliary axes, brings to
the effectively 2D case, with the following two-state matrices
\bea{\mathcal{R}}_{jk}=tr_i R_{ijk} \quad {or}\quad {\mathcal{R}}_{ik}=tr_j
R_{ijk}.\ena
 It means, that the relations (\ref{ttun}) at ${n_y}=1$ or
${n_x}=1$ ensure the integrability of the
corresponding 2D models with the matrices ${\mathcal{R}}_{ik}$ or
${\mathcal{R}}_{jk}$.

\subsection{Matrices with permutation operators: constant solutions.}

For 2D case we usually fix the initial condition, which exists for the most investigated models
and has physical background:
$
R_{ij}(0)=P_{ij}%
$.
 Here $P_{ij}:V_i\otimes V_j\to V_j\otimes V_i$, is the permutation operator, and this means that $\tau(0)$ is expressed by a shift operator, which ensures the locality of the spin model corresponding to the given $R$-matrix.
  Also we would like to recall that always it is possible to perform some symmetry transformations and re-normalization over the $R$-matrices which will let the relation (\ref{ttu}) unchanged. Particularly, $R(u)\to f(u)R(u)$, $R(u)\to R(a_0 u)$ transformations are permissible with arbitrary constant $a_0$ and arbitrary function $f(u)$.

In the described formulation of the 3D vertex models, where  each
$R$-matrix has one
quantum state (attached to the vertical axe $z$) and  two
auxiliary states  (on two horizontal axes $x,\; y$), the permutation operators can be chosen in different
ways.   The requirement of the locality of the underlying 2D
quantum spin models means that there must be a point
$\mathbf{u}_0$ at which  $R$-matrix is a permutation operator in
the following sense $R_{i_1 i_2 i_3}^{j_1 j_2
j_3}(0)=P_{123}^{312}\equiv\delta_{i_1}^{j_3}\delta_{i_2}^{j_1}\delta_{i_3}^{j_2}$,
or  $R_{i_1 i_2 i_3}^{j_1 j_2
j_3}(0)=P_{123}^{231}\equiv\delta_{i_1}^{j_2}\delta_{i_2}^{j_3}\delta_{i_3}^{j_1}$.
Note that there are 3D solutions induced from 2D  solutions $R^s(u)$ of
YBEs, e.g., $R_{i_1 i_2 i_3}^{j_1 j_2
j_3}(u)=\delta_{i_1}^{j_2}{R^s}_{i_2 i_3}^{j_1 j_3}(u)$. These
matrices are the solutions to  STE, i.e. the combined 3D equations (\ref{ybed}),
 with  $2$D intertwiner matrices ${\bar{R}}^s(u,w)$ and $2$D permutation operators, $P_{12}$ . 
 The corresponding spin models are described by  1D quantum spin models situated
 on the parallel chains of  $\{x,y\}$-plane. Another factorized  solution
  is the following matrix - $R_{i_1 i_2 i_3}^{j_1 j_2 j_3}(u)={R^s}_{i_1 i_2}^{j_1 j_2}\delta_{i_3}^{j_3}$,
   with the intertwiner operators equal to permutation matrices.
  We see, that this solution is not effective for finding $3$D solutions at all, as here $R^s$ acts only on the auxiliary spaces, by which the trace must be taken  in the definition of the transfer matrix.  Moreover, when trying to find the spectral parameter solutions,
  with  natural symmetries $R_{i_1 i_2 i_3}^{j_1 j_2 j_3}=R^{i_1 i_2 i_3}_{j_1 j_2 j_3}$, $R_{i_1 i_2 i_3}^{j_1 j_2 j_3}=R_{i'_1 i'_2 i'_3}^{j'_1 j'_2 j'_3}$, $i'=i+1\; {mod}\; 2$, we find that the only non-trivial solutions can be formulated by the mentioned effectively 2D solutions.

  The general 20-vertex model has
  the non-vanishing elements in the same positions of the matrix elements, as the
  matrix constructed only by means of the permutation operators and identity. It is the pure analogue of 2D
  case, as the six-vertex model's
   (1D XXZ spin-model relating to the quantum $SU_q(2)$ symmetry) trigonometric solution is the
  generalization of the rational solution $R=I+(1/u) P$
  (1D XXX-model, with non-deformed $SU(2)$ symmetry). 
%
%
  So, the simplest extension of the soluble 2D spin-models to 3D case
 could be in the form of the linear superposition:
 \bea R^p(\mathbf{u})=\sum_{n=1}^6 f_n(\mathbf{u}) P^{n}, \ena
  where we denote by $P^n$
  the permutation operators $P_{123}^{\{r_1 r_2
 r_3\}}$ with the matrix elements
 $\delta_{i_1}^{j_{r_1}}\delta_{i_2}^{j_{r_2}}\delta_{i_3}^{j_{r_3}}$. Here the indexes $\{r_1 r_2
 r_3\}$ are the possible permutations of the sequence $\{1,2,3\}$, including the identity
 (corresponding to the unity operator). Note that
$P_{123}^{231}=(P_{123}^{312})^2$,
$(P_{123}^{312})^3=P_{123}^{312} P_{123}^{231}=I$. Besides of full permutations, the next three
 permutations operators $P_{123}^{132}$, $P_{123}^{213}$, $P_{123}^{321}$, have the property $P^2=I$. Here also, if we look for
 the solutions, which at the fixed point equal to the full permutations
 $P_{123}^{312}$ or $P_{123}^{231}$ (locality principle), we shall find only rigid
 solutions. Next we demonstrate this for some examples.

 Let us particularly  consider the matrix with full
 permutation
  \bea& R(u)=f_i(u)I+f_p(u)P_{123}^{231}, &\ena
   by means of the stepwise consideration of the commutation relations (\ref{ttun}). We find that at $n_x=1$ or $n_y=1$ the relations are satisfied automatically. However the first non-trivial case at $n_x=2,\;n_y=2$:
   \bea&[\tau_{2,2}(u),\tau_{2,2}(v)]=8f_p^2(u)f_p^2(v)f_i(u)f_i(v)(f_p(u)f_i(v)-f_p(v)f_i(u))\Theta_{const},&\ena
   where $\Theta_{const}$
 is a constant $2^4\times 2^4$  matrix, brings to the relation $f_p(u)/f_i(u)=const$, i.e. to the rigid solution.  Another considering examples are the following choices of the matrices:
 $R(u)=f_i(u)I+f_p(u)P_{123}^{231}+f_r(u) P_{123}^{312}$ or $R(u)=f_i(u)I+f_p(u)P_{123}^{213}+f_r(u) P_{123}^{321}$.  At $n_x=2,\; n_y=2$ one can find  only one quadratic relation between the functions $f_{i,p,r}$, but  the next steps at $n_x=2,\;n_y=3$ and $n_y=2,\; n_x=3$ impose new additional constraints  on the coefficient functions, and the solutions become constant ones.  One can verify that these rigid matrices satisfy   local ZTE  or STE with appropriate constant intertwiners.

\subsection{Some simple 3D integrable models $R$-matrices}

   We construct at first a rather  trivial model, as an
   extension of the following 2D three-parametric matrix:
\bea
R(f_1,f_2,f_3)=
{\scriptsize{{\left(\ba{cccc}1&&&\\&&f_1&\\&f_2&&\\&&&f_3\ea
\right)}}}.\ena
This is obviously   solution to YBE
 $\bar{R}_{12}(f_i,g_i)R_{13}(f_i)R_{23}(g_i)=R_{23}(g_i)R_{13}(f_i)\bar{R}_{12}(f_i,g_i)$,
 $i=1,2,3$.
 Here one can fix $f_i=e^{u_i}$, if the system of YBE  is homogeneous
(intertwiner matrices $\bar{R}$ coincide with the quantum R-matrices),
and the spectral parameters are additive:
$R_{12}(u_i-v_i)R_{13}(u_i)R_{23}(v_i)=R_{23}(v_i)R_{13}(u_i)R_{12}(u_i-v_i)$.
The corresponding Hamiltonian operator, constructed in the ABA
 framework, reads in terms of the Pauli spin-$1/2$ operators as follows: $H=\sum_i(\alpha_1 S_z(i)\otimes
I(i+1))$ $+\alpha_2 I(i)\otimes S_z(i+1)$ $+\alpha_3 S_z(i)\otimes
S_z(i+1)$. Here we  take
$u_i=\alpha'_i u$, and  the constants $\alpha_i$ are connecting with
$\alpha'_i$ by linear relations. If $u_3=u_1+u_2$, then
this Hamiltonian describes a free-fermionic model (as the Ising,
$XX$ or $XZ$ models). In general, in terms of scalar fermions
 obtained by means of Jordan-Wigner transformation \cite{JW},
this is a Hamiltonian of this form 
$H=\sum_i(J_1 n_i+J_2 n_{i+1}+J_0 n_i
n_{i+1})=(J_1+J_2)\sum_i(n_i)+J_0\sum_i n_i n_{i+1}$. Here $n=(1-2s_z)/2$.  For 3D
model, we can present an analogous  solution, taking one of
two R-matrices $R^f$ and $P R^f$, with $P$ being a full
permutation operator, and with the following matrix
$R^f(f_1,f_2,f_3,f_4)$ which has
 eight non-zero elements and four arbitrary functions (spectral or colored parameters) $f_i$.
\begin{small}
\bea \label{r3-f}R^f(f_1,f_2,f_3,f_4)=
\scriptsize{\left(\ba{cccccccc}1&&&&&&&\\
&&&&f_1&&&\\
&f_2&&&&&&\\
&&&&&f_3&&\\
&&f_4&&&&&\\
&&&&&&\alpha_{14} f_1 f_4&\\
&&&\alpha_{24} f_2 f_4&&&&\\&&&&&&&\alpha_{34} f_3 f_4\ea\right),}
\ena\end{small}
%
with constants $\alpha$. The corresponding transfer matrices, constructed by means of the quantum $R^f$-matrices,  which have different sets of the functions $f_i$ (or spectral parameters), are commuting - $\left[tr[\prod R^f],tr[\prod
R^g]\right]=0$. 
 %
%
 The quantum $R^f$-matrices satisfy  ZTE-kind non-homogeneous
relations,
\bea \bar{R}^e \bar{R}^h R^{f} R^{g}=R^{g} R^{f} \bar{R}^h
\bar{R}^e \label{rfg} \ena
where the intertwiner matrices  $\bar{R}$ are invertible operators and  have a
bit different structure, also having eight non-zero elements (in the positions of the non-zero matrix elements of $\approx P_{13} R^f$), which are the rational functions over the functions $f_i/g_j$.
The combined equations  with  appropriate two dimensional intertwiner operators (i.e. STE)  also take place.
 And the intertwiner matrices, satisfying STE with the mentioned $R^f$ quantum
matrices, are the followings:
{
 \bea
 {R^e(f_i,g_i)=\scriptsize{
\left(\ba{cccc}1&&&\\
&&\frac{f_1}{g_1}&\\
&\frac{f_2 f_4}{g_2 g_4}&&\\
&&&\frac{f_3  f_4}{g_3 g_4}\ea\right)},\qquad
R^h(f_i,g_i)=
\scriptsize{\left(\ba{cccc}1&&&\\
&&\frac{f_1 f_4}{g_1 g_4}&\\
&\frac{f_2}{g_2}&&\\
&&&\frac{f_3 f_4}{g_3 g_4}\ea\right)}}
 \ena}
 However, if
 to demand, that there is a point ${\mathbf u_0}$, such that $R^f({\mathbf u_0})=P$, i.e. $a_{ij}=1,\; f_i=e^{a_i u},\; \{{\mathbf u_0}\}=\{0,0,0,0\},$ then the cell Hamiltonian
 of the corresponding 2D quantum chain model will have such elements $H_{i,j}\approx n_{i,j}+n_{i,j}n_{i+1,j+1}$. This means effectively interactions only in one direction. 
 However we can try to find the solutions with richer structure:
%
\begin{small}
\bea \bar{R}^f(f_1,f_2,f_3)=
\scriptsize{\left(\ba{cccccccc}1&&&&&&&\\
&f_1&a_1&&&&&\\
&f_2&&&&&&\\
&&&a_2 f_1&&a_2 f_2 f_3&&\\
&&\frac{a_4 f_1}{f_2 f_3}&&a_5&&&\\
&&&a_6 f_1&&\frac{a_7 f_1}{f_3}&\\
&&&\frac{a_8 f_1}{f_2}&&&a_9&\\&&&&&&&a_{10} f_1\ea\right),}
\ena\end{small}
Here there are ten constants $a_i$ and  three arbitrary functions.
If to check the local equations, we can find the appropriate non-trivial two-dimensional
intertwiner matrices:
\begin{small}
 \bea
 {\bar{R}^e(f_i,g_i)=\scriptsize{
\left(\ba{cccc}1&&&\\
&&\frac{f_2}{g_2}&\\
&\frac{f_1 g_2}{f_2 g_1}&&\\
&&&\frac{f_1}{g_1}\ea\right)},\qquad
\bar{R}^h(f_i,g_i)=\scriptsize{
\left(\ba{cccc}1&&&\\
&&\frac{f_1 g_3}{g_1 f_3}&\\
&\frac{f_2}{g_2}&&\\
&&&\frac{f_1}{g_1}\ea\right)}}
 \ena\end{small}
In the same way, we can present other series of the solutions, which have non-zero matrix elements
in the same positions as the $R$-matrix constructed by the permutation
operators and unit operator.  The part of such matrices it is possible
to present as factorizable operator, with corresponding 2D quantum
models, which contain interactions only along one direction, i.e. these are
actually one-dimensional interacting quantum models and all these matrices  may have their analogs
in the variety of the YBE solutions. However it is possible to construct
integrable even rather simple and essentially two-dimensional quantum
models by the three-dimensional R-matrices satisfying to transfer
matrix commutativity,  if we do not require the
locality of the 2D Hamiltonian operators or do not apply strong requirements on the form of R. In that cases there is a rich variety of
the solutions.
One can present numerous rational solutions from the transfer matrices commutativity $tr[t(f_i),t(g_i)]=0$, which will have non-local
 nature. An example of the simple solution, which is non-factorizable and
describes a non-local model with interaction in terms of the JW fermions, is the following two-spectral parametric $8\times 8$ matrix $r(f_1,f_2)$, with non-zero matrix elements $r_{ab}$, $a,b=1,...,8$,
\bea\nn & r_{ii}=\rho_i,\quad \{i=1,...,8\},\quad
 \quad r_{25}=f_1,\quad
r_{32}=f_2,&\\&
 r_{16}=\varepsilon_1 f_1,\quad
r_{38}=\varepsilon_2 f_1,\quad
r_{46}=\varepsilon_3 f_1 f_2,&\\& r_{53}= \frac{\varepsilon_4}{f_1 f_2},
\quad r_{67}=\frac{\varepsilon_5}{f_2},\quad
r_{74}=\frac{\varepsilon_6}{f_1},\quad r_{83}=\frac{\varepsilon_7}{f_1 }.&\nn\ena
Here the parameters $\rho_i, \;\varepsilon_i$ are constants, $f_i$ - the spectral parameters.  One can check the commutativity of the transfer matrices  using the 3D intertwiner matrices with similar structure. As well, the  STEs are satisfied, with the following 2D intertwiners ${\bar{r}^e},\;{\bar{r}^h}$,  $\bar{r}^e(f_i,g_i)\bar{r}^h(f_i,g_i)r(f_1,f_2)r(g_1,g_2)=r(g_1,g_2)r(f_1,f_2)\bar{r}^h(f_i,g_i)\bar{r}^e(f_i,g_i)$, having non-trivial elements $\bar{r}^{e,h}_{23}$ and $\bar{r}^{e,h}_{32}$:
\begin{small}
\bea
 \bar{r}^e(f_i,g_i)=\scriptsize{
\left(\ba{cccc}1&&&\\
&&\frac{f_1}{g_1}&\\
&\frac{g_1}{f_1}&&\\
&&&1\ea\right)},\qquad
\bar{r}^h(f_i,g_i)=
\scriptsize{\left(\ba{cccc}1&&&\\
&&\frac{g_2}{f_2}&\\
&\frac{f_2}{g_2}&&\\
&&&1\ea\right)}.\ena\end{small}%
%
\subsection{Perspectives: multi-qubit states, braiding transformations}

  In quantum information theory the unitary solutions to YBE attract an interest as candidates of the
   universal quantum gates, \cite{LK,ZKG}. The unitary braiding operation can entangle the full unentangled
   states, and as well,  the  unitary solutions to YBE, related with the braiding groups, can produce quantum gates. The constant solutions (as the simplest constant YBE solutions - the general inhomogeneous permutation operators), along with the Yang-Baxterized solutions have been  used to construct quantum gates. The spectral parameter defines the degree of the entanglement. In this context, we can
   consider also the solutions to 3D integrability equations. We see, that in the construction of the multi-qubit cluster states (see as example \cite{CXG,RB,PWT}), one use the products of YBE-solutions. Acting on the purely
    separable $N$-qubit state by the operators $R_{12}R_{23}...R_{N-1N}$ one can obtain N-qubit  Greenberger-Horne-Zeiliger (GHZ) states. Particularly, for three-qubit  cluster states the used operators $R_{12}R_{23}$-operators (or $R_{12}R_{23}R_{13}$) themselves can be the solutions to ZTE or STE (factorised solutions,  which we have called as inherited from the YBE-solutions). And one can try to involve in the construction of the full or mixed entangled  multi-qubit states, the constant or spectral-parameter  dependent  solutions to integrability equations for higher dimensions. As example, the action of the simple three-state $R$ operator (\ref{r3-f}), with the elements
    $\alpha_{ij}=e^{i \xi_{ij}}$, $f_i=e^{i \phi_i u}$ ($\xi_{ij}$, $\phi_i$ - constants), on the initial separable state
    $$\frac{(|0\rangle+|1\rangle)}{\sqrt{2}}\otimes \frac{(|0\rangle+|1\rangle)}{\sqrt{2}}\otimes \frac{(|0\rangle+|1\rangle)}{\sqrt{2}}$$
 brings to the following entangled state:
 \bea \nn \frac{|0\rangle\otimes |0\rangle\otimes |0\rangle+
 e^{i\phi_1 u}|1\rangle\otimes |0\rangle\otimes |0\rangle+
 e^{i \phi_2 u}|0\rangle\otimes |0\rangle\otimes |1\rangle}{2^{3/2}}+\\
 \frac{e^{i\phi_3 u}|1\rangle\otimes |0\rangle\otimes |1\rangle+
e^{i\phi_4 u}|0\rangle\otimes |1\rangle\otimes |0\rangle+e^{i\xi_{14}+i(\phi_1+\phi_4)u}
 |1\rangle\otimes |1\rangle\otimes |0\rangle}{2^{3/2}}+\\
 \frac{+e^{i\xi_{24}+i(\phi_2+\phi_4)u}
 |0\rangle\otimes |1\rangle\otimes |1\rangle+
 e^{i\xi_{34}+i(\phi_3+\phi_4)u}|1\rangle\otimes |1\rangle\otimes |1\rangle}{2^{3/2}}.\ena
 The quantities $|e^{i\xi_{24}}-1|$, $|e^{i\xi_{34}}- e^{i\xi_{14}}|$ and $|e^{i\xi_{14}}-1|$ measure the degree of the entanglement of the resulting state. If they all vanish, then the state factorises into the unentangled product of the separate states.

   In this work we have investigated the spectral-parameter dependent $20$-vertex models, which can induce 2D integrable lattice models, and  we do not give much attention to the constant solutions.  However, as it was stated above, there are many  constant (unitary) solutions to ZTE or STE. And the solutions with appropriate braiding properties can be as candidates for three-qubit quantum gates \cite{EDZ}. By means of the Scr\"{o}dinger equation from the evolution operator $R$ the corresponding Hamiltonian operators for the multi-qubit states (or linear clusters) are constructed  \cite{LK,ZKG,CXG}. The $R$-matrices' dependence from the time parameter in the  quantum information theory may differ from the usual formulation in the ABA. And now the usual spectral parameters can be considered as time-independent parameters (contributing to the entanglement), meanwhile the constants can be time-dependent, thus ensuring the entangled eigen-states for the resulting Hamiltonian operators \cite{CXG}. As an extension, the solutions to higher dimensional restricted equations can be considered in the context of multi-qubit quantum gates.

\section{The combined 3D equations in IRC  formulation}
\setcounter{equation}{0}

We also  can propose the
combined versions of the equations for the interaction-round-cube version \cite{BBt1} of ZTE (the analogue of the face-type YBE \cite{YPB,GP}), formulating as follows (\ref{W}),  see Fig. 2.

{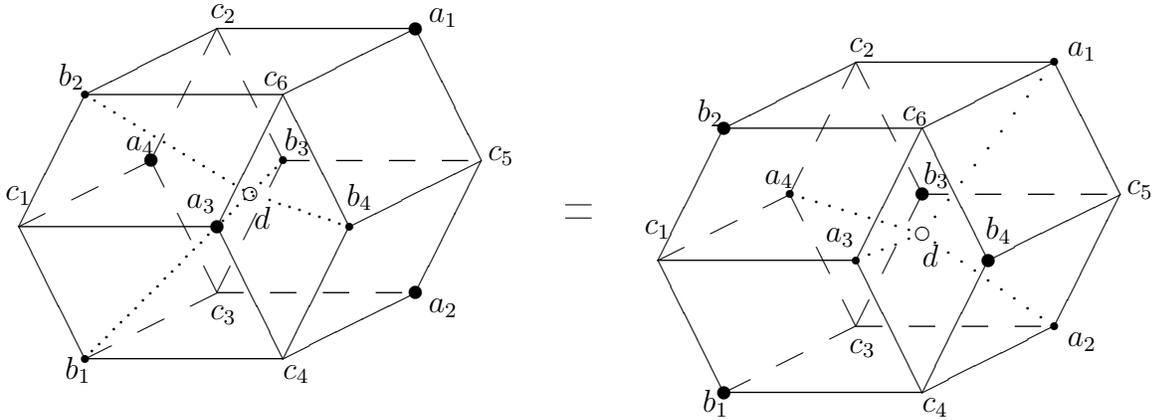
\begin{figure}[h]
\hspace{2cm} \unitlength=5pt
\begin{picture}(25,35)(12.5,0)
 \linethickness{0.5pt}
 \multiput(5,0)(15,0){2}{\line(-1,2){5}}
\multiput(5,0)(0,20){2}{\line(1,0){15}}
 \multiput(0,10)(15,0){2}{\line(1,2){5}}
\multiput(20,0)(10,5){2}{\line(1,2){5}}
 \multiput(0,10)(15,15){2}{\line(1,0){15}}
\multiput(25,10)(10,5){2}{\line(-1,2){5}}
\multiput(20,0)(5,10){2}{\line(2,1){10}}
\multiput(5,20)(15,0){2}{\line(2,1){10}}
\multiput(10,15)(5,-10){2}{\multiput(0,0)(2,4){3}{\line(1,2){1}}}
\multiput(5,0)(-5,10){2}{\multiput(0,0)(4,2){3}{\line(2,1){2}}}
\multiput(15,5)(5,10){2}{\multiput(0,0)(-2,4){3}{\line(-1,2){1}}}
\multiput(20,15)(-5,-10){2}{\multiput(0,0)(4,0){4}{\line(1,0){2}}}
\put(15,10){\circle*{1}} \put(30,5){\circle*{1}}
\put(30,25){\circle*{1}} \put(10,15){\circle*{1}}
\put(3.5,-1.5){$b_1$} \put(31,25.5){$a_1$} \put(31,3.5){$a_2$}
\put(8.1,15.8){$a_4$} \put(20.1,15.8){$b_3$} \put(35.5,15){$c_5$}
\put(-1,12){$c_1$} \put(3,20.7){$b_2$} \put(18.5,20.7){$c_6$}
\put(14.5,25.8){$c_2$}
\put(14.5,3){$c_3$}\put(20,-1.5){$c_4$}\put(12.7,11.4){$a_3$}
\put(24.8,11.6){$b_4$}\put(17.8,9.85){$d$}
 {\put(17.5,12.5){\circle{1}}}
 \put(25,10){\circle*{0.5}} \put(5,0){\circle*{0.5}}
\put(5,20){\circle*{0.5}} \put(20,15){\circle*{0.5}}
\multiput(5,20)(0.75,-0.45){18}{\circle*{0.1}}
\multiput(25,10)(-0.75,0.25){9}{\circle*{0.1}}
\multiput(20,15)(-0.56,-0.56){4}{\circle*{0.1}}
\multiput(5,0)(0.55,0.55){24}{\circle*{0.1}}
\end{picture}

\begin{picture}(5,2)(-42,0)
\multiput(-1,15)(0,1){2}{\line(1,0){2}}
\end{picture}

\begin{picture}(25,35)(-48,-37)
 \linethickness{0.5pt}
 \multiput(5,0)(15,0){2}{\line(-1,2){5}}
\multiput(5,0)(0,20){2}{\line(1,0){15}}
 \multiput(0,10)(15,0){2}{\line(1,2){5}}
\multiput(20,0)(10,5){2}{\line(1,2){5}}
 \multiput(0,10)(15,15){2}{\line(1,0){15}}
\multiput(25,10)(10,5){2}{\line(-1,2){5}}
\multiput(20,0)(5,10){2}{\line(2,1){10}}
\multiput(5,20)(15,0){2}{\line(2,1){10}}
\multiput(10,15)(5,-10){2}{\multiput(0,0)(2,4){3}{\line(1,2){1}}}
\multiput(5,0)(-5,10){2}{\multiput(0,0)(4,2){3}{\line(2,1){2}}}
\multiput(15,5)(5,10){2}{\multiput(0,0)(-2,4){3}{\line(-1,2){1}}}
\multiput(20,15)(-5,-10){2}{\multiput(0,0)(4,0){4}{\line(1,0){2}}}
\put(25,10){\circle*{1}} \put(5,0){\circle*{1}}
\put(5,20){\circle*{1}} \put(20,15){\circle*{1}}
\put(20,12){\circle{1}}
 \put(15,10){\circle*{0.5}}
\put(30,5){\circle*{0.5}} \put(30,25){\circle*{0.5}}
\put(10,15){\circle*{0.5}}
\multiput(10,15)(1,-0.3){10}{\circle*{0.1}}
\multiput(30,25)(-1,-1.3){10}{\circle*{0.1}}
\multiput(15,10)(1,0.4){5}{\circle*{0.1}}
\multiput(30,5)(-1,0.7){10}{\circle*{0.1}} \put(31,25.5){$a_1$}
\put(31,3.5){$a_2$} \put(8.1,15.8){$a_4$} \put(20.1,15.8){$b_3$}
\put(35.5,15){$c_5$} \put(-1,12){$c_1$} \put(3,20.5){$b_2$}
\put(18.5,20.6){$c_6$} \put(14.5,25.8){$c_2$}\put(3.4,-1.5){$b_1$}
\put(14.5,3){$c_3$}\put(20,-1.5){$c_4$}\put(12.7,11.4){$a_3$}
\put(24.8,11.6){$b_4$}\put(20,9.5){$d$}
\end{picture}\label{12edron}
\vspace{-6cm} \caption{IRC or 12-hedron /rhombic dodecahedron/ equation}
\end{figure}}

\unitlength=10pt

\newsavebox{\blokc}
\sbox{\blokc}{\begin{picture}(15,15)
\put(7.5,2.9){\line(0,1){9.2}} \put(2.5,7.5){\line(1,0){10}}
\put(5,5){\line(1,1){5}}
\end{picture}}

\newsavebox{\blokh}
\sbox{\blokh}{\begin{picture}(20,10)
\multiput(0,0)(5,5){3}{\multiput(0,0)(10,0){5}{\usebox{\blokc}}}
\end{picture}}


\newsavebox{\blokcc}
\sbox{\blokcc}{\begin{picture}(18,14)
\put(9,2.5){\line(0,1){9.2}} \put(4.5,7){\line(1,0){9.5}}
\put(5.5,5){\line(2,1){8}}
\end{picture}}

\newsavebox{\blokhh}
\sbox{\blokhh}{\begin{picture}(100,22)
\multiput(0,0)(8,4){3}{\multiput(0,0)(10,0){2}{\usebox{\blokcc}}}
\end{picture}}

\bea\nn &\sum_d
W(a_4|c_2,c_1,c_3|b_1,b_3,b_2|d)W'(c_1|b_2,a_3,b_1|c_4,d,c_6|b_4)W''(b_1|d,c_4,c_3|a_2,b_3,b_4|c_5)&\\\nn&
W'''(d|b_2,b_4,b_3|c_5,c_2,c_6|a_1)=\sum_d
W'''(b_1|c_1,c_4,c_3|a_2,a_4,a_3|
d)\times &\\\label{W}& W''(c_1|b_2,a_3,a_4|d,c_2,c_6|a_1)
W'(a_4|c_1,d,c_3|a_2,b_3,a_1|c_5)W(d|a_1,a_3,a_2|c_4,c_5,c_6|b_4),&\\
 \nn & \sum_d
W(a|b_1,b_2,b_3|c_1,c_2,c_3|d)W^{-1}(d|b'_1,b'_2,b'_3|c'_1,c'_2,c'_3|a')=
\delta_{a}^{a'}\delta_{b_1}^{b_1'}\delta_{b_2}^{b_2'}\delta_{b_3}^{b_3'}
\delta_{c_1}^{c_1'}\delta_{c_2}^{c_2'}\delta_{c_3}^{c_3'},&\ena

We suggest a version of the combined IRC in the
following form, where the intertwiner matrices $W$, $\bar{W}$ and $W'$ now are 2D
face matrices, with four indexes,
\bea\nn \sum_d
W(a_4|c_2|_c3|b_3)\bar{W}(c_1|b_2|b_1|d)W'(a_3|c_6|c_4|b_4)\\\nn W''(b_1|d,c_4,c_3|a_2,b_3,b_4|c_5)
W'''(d|b_2,b_4,b_3|c_5,c_2,c_6|a_1)=\\
\nn\sum_d
W'''(b_1|c_1,c_4,c_3|a_2,a_4,a_3|
d)W''(c_1|b_2,a_3,a_4|d,c_2,c_6|a_1)\\\label{Ww}
W'(a_4|c_2|_c3|b_3)\bar{W}(d|a_1|a_2|c_5)W(a_3|c_6|c_4|b_4). \ena
These equations also are sufficient for the transfer matrix commutativity in the same foot as the ordinary 12-hedron  equations (the detailed  proof for vertex version see in \cite{ASHS}).

As for the ordinary vertex models, here also one can find out some immediate solutions - at least the factorised solutions constructed by means of 2D face-YBE solutions. They can be the analogs of the corresponding STE solutions, taking into account the vertex-face correspondence brought in Fig. \ref{Figv-f} (which is not  universal correspondence, however).

We here propose also another combined  IRC  3D equations, which  can be projected onto the
face-YBEs in  2D space of the vertexes
$\{c_1|a_3,a_4|d|b_3,b_4|c_5\}$:
\bea\nn \sum_d
W(a_4|c_1|b_3|d)W'(c_1|a_3|d|b_4)W''(b_1|d,c_4,c_3|a_2,b_3,b_4|c_5)\times\\\nn
W'''(d|b_2,b_4,b_3|c_5,c_2,c_6|a_1)=\sum_d
W'''(b_1|c_1,c_4,c_3|a_2,a_4,a_3|
d)\times\\\label{Www}
W''(c_1|b_2,a_3,a_4|d,c_2,c_6|a_1)
W'(a_4|d|b_3|c_5)W(d|a_3|c_5|b_4). \ena
The graphical representations of these two versions of the combined IRC equations (\ref{Ww}) and (\ref{Www}) can be obtained from Fig. 2 easily, following to the indexes of  the operators in the equations.

\section{The  combined cube  equations in semi-check
formalism:  non-local 2D quantum 
integrable models}
\setcounter{equation}{0}%

The cube
equations, suggested in \cite{KhFKS}, contain four 3D $R$-matrices defined on the cubes, as IRC equations. However the cube equations by their operator form are more similar to non-symmetric vertex-like equations, as the four $R$-matrices
 have not equivalent roles in the equations. Two quantum $R$-matrices, by which the transfer matrices are constructed, and
  as well as two intertwiner matrices,  have two common vertexes and one common link, meanwhile each pair of quantum and intertwiner matrices have only one common vertex (see the figures in \cite{KhFKS}).
 The graphical structure of the corresponding operators, acting on the
 tensor product of  four vector spaces, situated on the lattice vertices %
    is
presented in Fig. 1 by $R_{afgb}^{edch}$. The cube equations,
  (Fig. 3 of the paper \cite{KhFKS}), are suitable for the models
with the checkerboard-like  Hamiltonians, for which transfer
matrices are constructed by means of the product of two 2D  transfer
matrices. And hence, two sets of the cube equations must be considered. Here, in this
article, the equations we present in the  semi-check
formulation, after acting by a permutation operator on the
$R$-matrices, $P R=\breve{R}$. And correspondingly, the indexing of the equations are
changed in such a way, that  only one set
of the cube equations is sufficient to ensure the integrability.  And the definition of the corresponding 3D statistical models must be slightly changed conditioned by the shift of the neighboring monodromy matrices in the partition function.
 The semi-check matrices we  define in the following
 way - $\breve{R}_{afgb}^{edch}=R_{afgb}^{ched}$, with  appropriate chosen permutations. For the quantum matrices the first two pairs of the upper and lower  indexes we can refer as the quantum states' indexes,
 the next two - as the auxiliary states indexes. The direct construction of
the  cube equations with two kind of quantum states for these
 matrices,
$\breve{R}_{\alpha\beta\gamma\delta}^{\delta'\gamma'\beta'\alpha'}$
 will look like as
\bea &&\breve{R^1}_{i_1 i_2 i_3 i_4}^{j_1 j_2 j_3
j_4}(u)\breve{R^2}_{i_5 i_6 j_3 j_4}^{j_5 j_6 k_3
k_4}(v)\breve{R^3}_{j_2 i_6 i_7 i_8}^{k_2 k_6 j_7
j_8}(w)\breve{R^4}_{j_1 j_5 j_7 j_8}^{k_1 k_5 k_7
k_8}(y)=\nonumber\\&&\breve{R^4}_{i_2 i_6 i_7 i_8}^{j_2 j_6 j_7
j_8}(y)\breve{R^3}_{i_1 i_5 j_7 i_8}^{j_1 j_5 k_7
j_8}(w)\breve{R^2}_{j_1 j_2 i_3 i_4}^{k_1 k_2 j_3
j_4}(v)\breve{R^1}_{j_5 j_6 j_3 j_4}^{k_5 k_6 k_3 k_4}(u).
\label{check}\ena
 The corresponding  combined equations with the restricted intertwiners are
\bea &&\breve{R^1}_{i_1 i_2 i_3 i_4}^{j_1 j_2 j_3
j_4}(u)\breve{R^2}_{i_5 i_6 j_3 j_4}^{j_5 j_6 k_3
k_4}(v)\breve{R^3}_{j_2 i_6}^{k_2 k_6}(w)\breve{R^4}_{j_1 j_5}^{k_1 k_5}(y)=\nonumber\\&&\breve{R^4}_{i_2 i_6 }^{j_2 j_6}(y)\breve{R^3}_{i_1 i_5}^{j_1 j_5}(w)\breve{R^2}_{j_1 j_2 i_3 i_4}^{k_1 k_2 j_3
j_4}(v)\breve{R^1}_{j_5 j_6 j_3 j_4}^{k_5 k_6 k_3 k_4}(u).
\label{check-r}\ena
It is remarkable that if to merge the first pair of the states of quantum and intertwiner $\breve{R}$-matrices into one state, then for the obtained three-state matrices from the Eqs. (\ref{check}, \ref{check-r}) one can recover the vertex version of ZTE and STE  as limited cases.  Taking the quantum matrices $\breve{R}(u)$ and $\breve{R}(v)$ as extension of  $XYZ$ model's 2D matrix, 
 which in operator form can be presented as
$\breve{R}_{1,2}=I\otimes I\otimes I\otimes I+\sum_{i=\{x,y,z\}}u_i(v_i)\sigma^{i}\otimes
\sigma^{i}\otimes \sigma^{i}\otimes \sigma^{i}$ (here $I$
is the unity matrix and $\sigma^{i}$-s are Pauli $2\times 2$
matrixes), then the intertwiner operators corresponding to the matrixes
$\breve{R}(w)$, $\breve{R}(y)$ in the equations (\ref{check}) are simply the
constant operators
$\breve{R}_{3,4}=I\otimes I\otimes I\otimes
I+\sigma^{z}\otimes \sigma^{z}\otimes \sigma^{z}\otimes
\sigma^{z}\pm \sigma^{x}\otimes \sigma^{x}\otimes
\sigma^{x}\otimes \sigma^{x}\mp \sigma^{y}\otimes
\sigma^{y}\otimes \sigma^{y}\otimes \sigma^{y}$ or
$\breve{R}_{3,4}=I\otimes I\otimes I\otimes I-\sigma^{z}\otimes
\sigma^{z}\otimes \sigma^{z}\otimes \sigma^{z}\pm
\sigma^{x}\otimes \sigma^{x}\otimes \sigma^{x}\otimes
\sigma^{x}\pm \sigma^{y}\otimes \sigma^{y}\otimes
\sigma^{y}\otimes \sigma^{y}$.
 Also one can find the intertwiner solutions to the combined cube equations (\ref{check-r}).
  In contrast to the ordinary cube equations (see \cite{KhFKS}), the solutions to
these semi-check equations and the corresponding 2D quantum
Hamiltonian operators, obtained in the expansion of the transfer matrices at $u=u_0$, supposing $\breve{R}_{1,2}(u_0)=I\otimes I\otimes I\otimes
I$, do not describe local models with nearest-neighbor
interactions. E.g, the obtained solutions describe long range interactions along one of the
axes: $H=\sum_{i,i',j}H_{\{i,j\},\{i',j+1\}}$.

\section{New extension of the star-triangle relations}
\setcounter{equation}{0}

 In the early works on  2D statistical models the star-triangle relations
   figurate as local integrability condition  \cite{Y}. In the cases, when the
     statistical sums  are expressed by the local weights defined on the links (graph models, Potts model), i.e.
     $Z=\sum_{configurations}\prod_{i,j}^{N_1,N_2}$ $\left[W^h(i,j;i+1,j)W^v(i,j;i,j+1)\right]$, then the star-triangle relations are defined as
     \bea
    \sum_d R_{pqr} \bar{W}_a^d(p,q) W_b^d(p,r) \bar{W}_d^c(q,r)=W_a^b(q,r) \bar{W}_a^c(p,r) W_b^c(p,q),\\
     W_c^a(q,r) \bar{W}_c^b(p,r) W_a^b(p,q) =\sum_d \bar{R}_{pqr}\bar{W}_c^d(p,q) W_d^a(p,r) \bar{W}_d^b(q,r).
     \ena
      Here $W_a^b(p,q)$ is the horizontal link weight $W^h(i,j;i+1,j)$, connecting the states denoted by $a,\; b$
      situated at the sites $(i,j)$ and $(i+1,j)$, and correspondingly, the vertical weights ${\bar{W}}_a^b(p,q)$ ($W^v(i,j;i,j+1)$) connect the states denoted by $a,\; b$
      situated at the sites $(i,j)$ and $(i,j+1)$. A pair of the spectral parameters - rapidities, directed along the
      links of the dual lattice, is attached to each link (see, e.g., \cite{YPBS,GSA,BBt1}).
      The factors $R_{pqr}$ do not depend from the state variables. In the  Fig. 3 one can find the graphical
 picture of the above equations, considering only the graphs with the vertices denoted by $\{a_1,\;b_1,\; c_1,\;d_1\}$ or $\{a_2,\;b_2,\; c_2,\;d_2\}$ (belonging to parallel planes).

         For the three-dimensional statistical models, defined on the regular cubic lattice, there is an additional weight
       ${\tilde{W}}_{a}^b=W^o(i,j,k;i,j,k+1)$, attached to the links 
        orthogonal  to  the horizontal and vertical links. We suggest here to extend the star-triangle relations into the following 3D form (Fig. 3):
 \bea
   R(p,q;\rho_p,\rho_q;r) W_{b_1}^{a_1}(p,q,\rho_q) {\bar{W}}_{b_1}^{c_1}(p,r,\rho_p) W_{a_1}^{c_1}(q,r,\rho_q) {\tilde{W}}_{a_1}^{a_2}(\rho_q,r)\times \\\nn {\tilde{W}}_{c_1}^{c_2}(\rho_p,r)W_{b_2}^{a_2}(p,q,\rho_q) \bar{W}_{b_2}^{c_2}(p,r,\rho_p) W_{a_2}^{c_2}(q,r,\rho_q ) =\\
    \nn  \sum_{d_1}\sum_{d_2} \bar{W}_{b_1}^{d_1}(q,r,\rho_q) W_{a_1}^{d_1}(p,r,\rho_p) \bar{W}_{d_1}^{c_1}(p,q,\rho_q)
    \tilde{W}_{a_1}^{a_2}(\rho_p,r)\times\\\nn \tilde{W}_{d_1}^{d_2}(\rho_q,r)
    \bar{W}_{b_2}^{d_2}(q,r,\rho_q) W_{a_2}^{d_2}(p,r,\rho_p) \bar{W}_{d_2}^{c_2}(p,q,\rho_q) ,\\
      \sum_{d_1}\sum_{d_2} {\bar{W}}_{c_1}^{d_1}(p,q,\rho_q) W_{b_1}^{d_1}(p,r\rho_p) {\bar{W}}_{d_1}^{a_1}(q,r,\rho_q)\times\\\nn
       {\tilde{W}}_{d_1}^{d_2}(\rho_q,r) {\tilde{W}}_{b_1}^{b_2}(\rho_p,r)
      {\bar{W}}_{c_2}^{d_2}(p,q,\rho_q) W_{b_2}^{d_2}(p,r,\rho_p) {\bar{W}}_{d_2}^{a_2}(q,r,\rho_q) =\\
      \nn {\bar{R}}(p,q;\rho_p,\rho_q;r) W_{b_1}^{c_1}(q,r,\rho_q) {\bar{W}}_{c_1}^{a_1}(p,r,\rho_p) W_{b_1}^{a_1}(p,q,\rho_q) \tilde{W}_{c_1}^{c_2}(\rho_p,r) \times\\\nn {\tilde{W}}_{b_1}^{b_2}{\rho_q,r}W_{b_2}^{c_2}(q,r,\rho_q) {\bar{W}}_{c_2}^{a_2}(p,r,\rho_p) W_{b_2}^{a_2}(p,q,\rho_q).
     \ena

\begin{figure}[t]
\unitlength=12pt
\begin{picture}(100,10)(5,-1)

\newsavebox{\rmatrixka}

\sbox{\rmatrixka}{\begin{picture}(0,0)(-2.5,5)
\put(3,4){\line(2,3){2}}\put(5,7){\line(-1,1){2}}
\put(6,6){\line(2,3){2}}
\put(5,7){\line(3,2){3}}
\put(3,9){\line(3,2){3}}
\put(8,9){\line(-1,1){2}}
\put(3,4){\line(0,1){5}}\put(6,6){\line(0,1){5}}
\multiput(3,4)(3,2){2}{\circle{0.3}}
\multiput(3,9)(3,2){2}{\circle{0.3}}
\multiput(5,7)(3,2){2}{\circle{0.3}}
\put(2,4){\scriptsize$b_1$}\put(2,9){\scriptsize$a_1$}\put(6.5,5.7){\scriptsize$b_2$}
\put(6.5,11){\scriptsize$a_2$}
\put(5,6,5){\scriptsize$c_1$}
\put(8.3,9){\scriptsize$c_2$}
\end{picture}}

\newsavebox{\rmatrixkb}

\sbox{\rmatrixkb}{\begin{picture}(0,0)(-2.5,5)
\put(3,4){\line(2,3){2}}\put(5,7){\line(-1,1){2}}
\put(6,6){\line(2,3){2}}
\put(5,7){\line(3,2){3}}
\put(5,7){\line(3,-1){3}}\put(8,9){\line(3,-1){3}}
\put(3,9){\line(3,2){3}}
\put(8,9){\line(-1,1){2}}
\multiput(3,4)(3,2){2}{\circle{0.3}}
\multiput(3,9)(3,2){2}{\circle{0.3}}
\multiput(5,7)(3,2){2}{\circle*{0.3}}
\multiput(8,6)(3,2){2}{\circle{0.3}}
\put(2,4){\scriptsize$b_1$}\put(2,9){\scriptsize$a_1$}\put(6.5,5.7){\scriptsize$b_2$}
\put(6.5,11){\scriptsize$a_2$}
\put(4,7){\scriptsize$d_1$}\put(8.3,9){\scriptsize$d_2$}
\put(8,5.2){\scriptsize$c_1$}
\put(11.1,7.2){\scriptsize$c_2$}
\end{picture}}

\put(1,0){\usebox{\rmatrixka}}\put(12.7,2.3){$=$}
\put(9.3,0){\usebox{\rmatrixkb}}
\put(11,-2){(i)}
\newsavebox{\rmatrixkc}

\sbox{\rmatrixkc}{\begin{picture}(0,0)(-2.5,5)
\put(1,6){\line(2,3){2}}\put(3,4){\line(-1,1){2}}
\put(4,8){\line(2,3){2}}\put(1,6){\line(3,2){3}}
\put(3,4){\line(3,2){3}}
\put(4,8){\line(1,-1){2}}
\put(1,6){\line(-3,1){3}}\put(4,8){\line(-3,1){3}}
\multiput(3,4)(3,2){2}{\circle{0.3}}
\multiput(3,9)(3,2){2}{\circle{0.3}}
\multiput(-2,7)(3,2){2}{\circle{0.3}}
\multiput(1,6)(3,2){2}{\circle*{0.3}}
\put(2,4){\scriptsize$b_1$}\put(2.1,9){\scriptsize$a_1$}\put(6.5,5.7){\scriptsize$b_2$}
\put(6.5,11){\scriptsize$a_2$}
\put(4.3,8){\scriptsize$d_2$}
\put(0.1,5.5){\scriptsize$d_1$}
\put(-2,6.1){\scriptsize$c_1$}
\put(1,8.1){\scriptsize$c_2$}
\end{picture}}

\newsavebox{\rmatrixkd}

\sbox{\rmatrixkd}{\begin{picture}(0,0)(-2.5,5)
\put(1,6){\line(2,3){2}}\put(3,4){\line(-1,1){2}}
\put(4,8){\line(2,3){2}}\put(1,6){\line(3,2){3}}
\put(3,4){\line(3,2){3}}
\put(4,8){\line(1,-1){2}}
\put(3,4){\line(0,1){5}}\put(6,6){\line(0,1){5}}
\multiput(3,4)(3,2){2}{\circle{0.3}}
\multiput(3,9)(3,2){2}{\circle{0.3}}
\multiput(1,6)(3,2){2}{\circle{0.3}}
\put(2,4){\scriptsize$b_1$}\put(2.1,9){\scriptsize$a_1$}\put(6.5,5.7){\scriptsize$b_2$}
\put(6.5,11){\scriptsize$a_2$}
\put(4.3,8){\scriptsize$c_2$}
\put(0.15,5.5){\scriptsize$c_1$}

\end{picture}}

\put(25.5,0){\usebox{\rmatrixkc}}
\put(33.5,0){\usebox{\rmatrixkd}}
\put(34.3,2.3){$=$}
\put(33,-2){(ii)}

\put(23,7){\vector(1,0){1}}\put(23.1,7.3){\scriptsize{p,q}}
\put(24.5,5.5){\vector(0,1){1}}\put(25,5.7){${\scriptsize{r}}$}
\put(25,7.3){\vector(3,2){1}}\put(25.4,7){${\scriptsize{\rho}}_{p,q}$}
\end{picture} \caption{Extension of the Star-triangle relations for 3D lattice: connected star-triangle relations - (i) and
   (ii)}\label{fig1}
\end{figure}
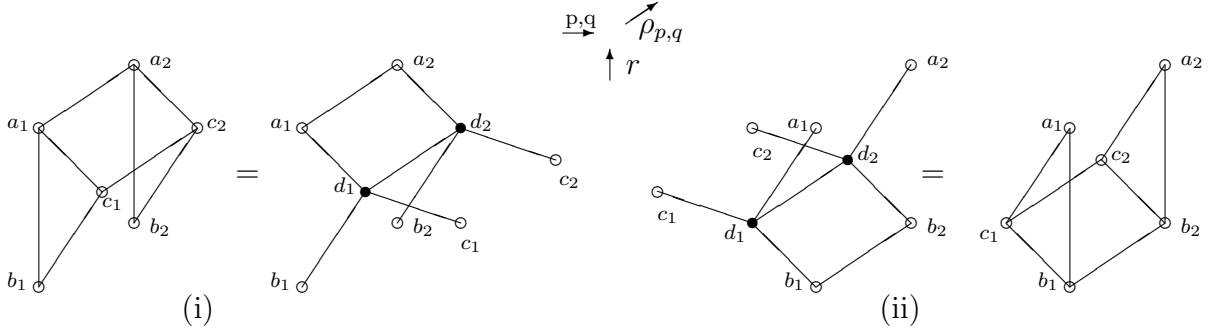
%
%
These equations are constructed in  such way, that their successive applications in the corresponding monodromy matrices' products, must take into account all the link weights only once. Formally they contain plaquette weights as products of the surrounding three link weights: say as $W_{a_1}^{c_1}(q,r,\rho_q) {\tilde{W}}_{a_1}^{a_2}(\rho_q,r) W_{a_2}^{c_2}(q,r,\rho_q )$ or $ W_{b_1}^{d_1}(p,r\rho_p) {\tilde{W}}_{b_1}^{b_2}(\rho_p,r) W_{b_2}^{d_2}(p,r,\rho_p)$.
Such generalization of the star-triangle relations one can refer either as combined  (relations which combine together the link and the {plaquette weights}) or preferably “connected” star-triangle relations, because of their form of two connected 2D star-triangle relations. Meanwhile this form of the equations in a sense can be associated also with the semi-check (\ref{check}) variation of the cube equations, in the same way, as  2D star-triangle equations are related to the face-type Yang-Baxter equations. In  Fig. 4 we are presenting the  fragments of the products of the transfer matrices and the {plaquette weights} figured in the
{“connected”} star-triangle relations. Note, that one can modify  by different variations  the {“connected”} star-triangle relations, changing the dispositions of the links in the construction of the  {plaquette}. The generalisation to the case of higher dimension $[d+1]$ also can be realised. In contrast to the the vertex case of the equations, when  in the generalisation to  combined equations at $d>2$ we add only  additional  auxiliary states,  here we shall add also the quantum states for having {“connected”}  $d$ star-triangle relations.

  In the “connected” equations we  attached third spectral (“rapidity”) parameter $\rho_{p}$
       to the weight links.  This “rapidity”  can be directed just along to the links (corresponding to $\tilde{W}$) being  orthogonal to the surface  characterized by the  “rapidities” $\{p/q,\;r\}$, which as usual are attached to the dual lattice for 2D \cite{BBt1}.
To each link now  three spectral parameters $\{p,\rho_p,r\}$ are attached, where the additional new third parameter $\rho$, in contrast to the situations  in two dimensional cases, can be associated with the links parallel to the horizontal ones, connected the vertices $e_1,\;e_2$, ($e=\{a,\;b,\;c,\;d\}$ in Fig. 3), i.e. along the links of the weights $\tilde{W}$. The parameters $p,\; r$ are attached to the weights $W,\; \bar{W}$ in standard way,
 supposing, that at  special values $\rho=\rho_0$, the extended equations can be reduced to the 2D  star-triangle relations. As  the projections of the suggested 3D equations turn into the 2D star-triangle equations, then it is natural to propose solutions as the generalizations of the known 2D solutions: $W^{s}_{3D}(p,q,\rho_0)=W^{s}_{2D}(p,q)$, where the weights $W^{s}_{2D}$ satisfy  the usual 2D  star-triangle  relations. Thus one can immediately present two rather trivial solutions  to 3D “connected” equations as natural
 extensions of $W^{s}_{2D},\; \bar{W}^{s}_{2D}$ (being the solutions to the 2D star-triangle relations):
 \bea&({ }^*)\:\tilde{W}_{e_1}^{e_2}(p,r,\rho_p)=constant,\; W(p,r,\rho_p)=W^{s}_{2D}(p,r),\; \bar{W}(p,r,\rho_p)=\bar{W}^{s}_{2D}(p,r),\; \rho_p\rho_0,&\nn\\
 &({ }^{**})\;\tilde{W}_{e_1}^{e_2}(p,r,\rho_p)={\delta}_{e_1}^{e_2},\; W(p,r,\rho_p)=\sqrt{W^{s}_{2D}(p,r)},\; \bar{W}(p,r,\rho_p)=\sqrt{\bar{W}^{s}_{2D}(p,r)},\; \rho_p=\rho_0.&\nn\ena
The presented  induced solutions  are boundary cases in the following sense: the first one $({ }^*)$ corresponds to the case, when the statistical sum can be factorised into the product of the statistical sums independent one from another defined on the parallel 2D planes; the second solution $({ }^{**})$ corresponds to the fully coupled case. The factors $R(p,q;\rho_p,\rho_q;r)$ also can be taken  as $ R^s_{2D}(p,q,r)$ (solutions to 2D relations).
In the work \cite{BBt1} the authors present a solution to the 3D (restricted) star-triangle relations, which are related to the IRC  equations \cite{BBt}, referred in the previous section. That solution coincides with the 2D Potts model's matrix elements (\cite{YPB,chBS}), with appropriate interpretation. They consider two type weights, which are attached to the links (interactions of two-vertexes) and to the triangles (three vertexes). In our case we have proposed another type of generalization of the star-triangle equations, with the weights defined only on the links connecting two vertexes.

 \section{Summary}
\setcounter{equation}{0}

 The investigation is devoted to the questions  on the integrability in the ABA technique appropriate for the statistical models in general high dimensions.   At
 first we constructed the so-called  combined integrability equations
  for general $\mathbb{D}$-dimensional  statistical models (\ref{ybe1}), which also can be considered as restricted or simplified $\mathbb{D}$-simplex equations, with intertwiner matrices of less dimensions than the quantum $R$-matrices. 
 Then, we  demonstrated  a rather general stepwise
 analysis (\ref{ttun}), for verifying whether  there  are commutative transfer matrices with the given fixed form of
 $R$-matrices. After definite number of steps, conditioned by the dimensions of the matrices and the symmetry properties, the analysis of the farther commutation relations (\ref{ttun}) may lead to two possible situation: the absence of the non-trivial solutions or repeating of the obtained constraints.
  In the second case  one can concretize the solutions for checking the local integrability equations to fix appropriate intertwiners,
  considering in this stage the inhomogeneous YBE (\ref{ybe1}).
  The examples are presented:  the general non-homogeneous eight vertex model has been revealed easily. For the case $\mathbb{D}=3$
  some simple spectral parameter dependent solutions also are obtained and the corresponding quantum 2D square lattice models are discussed.
    At the same time it is proved that there is no entirely new spectral parameter dependent  solutions constructed by means of the sum of 3D unity operator and the full projection operators. For more general $20$-vertex situation there are constant
solutions only and the solutions which are factorizing into  2D solutions of  YBE. However it is possible to obtain entirely 3D (not factorizing into 2D matrices) solutions with more general structure of the  vertex matrices: we constructed some simple examples (without locality property) in the  end part of Subsection 3.2. The Subsection 3.3 devotes to the another possible application of the discussed multi-particle matrices.  We propose, that the unitary solutions to ZTE (STE) are interesting also in the quantum information theory as candidates for the quantum three-qubit gates. And here the constant unitary  solutions, with appropriate braiding properties, have also an important role. The N-qubit gates (or multi-qubit linear cluster states) in the same way  can be considered among the solutions to the multi state integrability equations. Particularly, the $2^3\times 2^3$ three-qubit gate  $B_{12}B_{23}$ constructed in \cite{CXG}, which produces  three-qubit GHZ states by means of  $2^2\times 2^2$ braiding  $B$-matrices (yielding Bell states), satisfy the set of STE with two-dimensional intertwiner operators being usual permutation operators.

    In a sense, the presented stepwise scheme in Subsection 2.2
can be regarded as a test method. For small size lattices with powerful calculating techniques one can entirely solve the problem. In the case having positive
results (finding solutions for small $n$-s) one can then check the existence of  intertwiner operators using inhomogeneous  multidimensional combined equations.
  It would be more interesting to realize this concept in such cases,
when the direct consideration of the local integrability
conditions is not so effective.  In the work \cite{SHKAS-IP} we have constructed  3D $R$-matrix for the chiral Potts model on the regular cubic lattice. And it would be valuable to obtain the
    integrability's  possibility for this case just from the transfer matrices commutativity.
    And it seems from our preliminary calculations, that for these models there are no other integrability conditions,
     except of those
    which are connected to the 2D projections of the problem.

 And, the next issues investigated here  concern to the  3D analogues of the so-called
  face-type
  models, for which the states are attached to the vertexes. We proposed and  studied various combined versions   of the local compatibility conditions, which include simultaneously  2D two-state and 3D four-state $R$-matrices.
 And finally, we developed new analogue of the star-triangle equations for the case of 3D graph-type statistical  models.  The generalization to $D>3$ case can be done in the same manner.

\paragraph{Acknowledgments}
The work was supported by the Science Committee of RA, in the
frames of the research projects N 20TTWS-1C035, N 20TTAT-QTa009 and N 21AG-1C024.

\paragraph{Data Availability Statement} This manuscript has no associated data. 

\setcounter{section}{0}
\renewcommand{\thesection}{\Alph{section}}\Alph{section}
\section{Appendix}
\renewcommand{\theequation}{\Alph{section}.\arabic{equation}}
\addtocounter{section}{0}\setcounter{equation}{0}
  \setcounter{equation}{0}

{\begin{small}
  For the well known  eight-vertex models \cite{Baxt}  with the matrix elements
   $
   R_{ij}^{kr}(u),\quad i+j=k+r \; {mod}\; 2$,
the defining relations for the solutions following from the equations (\ref{ceq-n}) start from the low
dimensions $n=2,3$.  At $n=2$ there are two relations, which appear
to be sufficient for commutation of the transfer matrices
 (with two arbitrary constant parameters $d_0,\; a_0$)
\bea &\frac{R_{00}^{11}(u)R_{01}^{10}(u)}{R_{11}^{00}(u)R_{10}^{01}(u)}=d_0,
\label{t2h} \quad\frac{\left(
[R_{00}^{00}(u)]^2-[R_{11}^{11}(u)]^2+[R_{10}^{10}(u)]^2-[R_{01}^{01}(u)]^2\right)}{R_{11}^{00}(u)R_{01}^{10}(u)}=a_0.& \ena
Let us now separately discuss the situations with symmetric (*) and non-symmetric matrices (**).
\paragraph{* Symmetric $R$-matrices.} The following symmetry relations  $R_{00}^{00}(u)={R_{11}^{11}}(u)$, $R_{01}^{01}(u)={R_{10}^{10}}(u)$ and $R_{01}^{10}(u)={R_{10}^{01}}(u)$, $R_{00}^{11}(u)=d_0 {R_{11}^{00}}(u)$,
imply that the equations (\ref{t2h}) take place  automatically. Now $a_0=0$, and the  equations in (\ref{t2h}) are identities. One can always take $d_0=1$, as all the discussed equations (the commutativity of the transfer matrices, the ordinary and nonhomogeneous  Yang-Baxter equations) are defined up to the transformations: $R_{00}^{11}(u)\to R_{00}^{11}(u)/\sqrt{d_0}$, ${R_{11}^{00}}(u)\to \sqrt{d_0} {R_{11}^{00}}(u)$.

At the next steps, when $n=3$ and $n=4$ there are arisen the following constraints
correspondingly
{\bea
&\frac{(R_{00}^{00}(u)-R_{01}^{01}(u))^2-R_{01}^{10}(u)^2-d_0 {R_{11}^{00}(u)}^2}{R_{11}^{00}(u)}={\mathrm{const}},\quad{\mathrm{and}}\quad
\frac{R_{00}^{00}(u)R_{01}^{01}(u)}{R_{11}^{00}(u)}={\mathrm{const}}.\label{s12}&
\ena}%

\paragraph{** Non-symmetric matrices.} Here $a_0\neq 0$
 in (\ref{t2h}) and
at the next step $n=3$  the defining relations are the following ones.
\bea\label{nsym-r}
&\frac{R_{00}^{00}(u)([R_{00}^{00}(u)]^2-[R_{01}^{01}(u)]^2)-R_{10}^{10}(u)([R_{11}^{11}(u)]^2-[R_{10}^{10}(u)]^2)}{
(R_{11}^{11}(u)+R_{01}^{01}(u))( R_{01}^{10}(u)
R_{10}^{01}(u)+R_{00}^{11}(u)R_{11}^{00}(u))+(d_{+})R_{11}^{00}(u)R_{01}^{10}(u)(R_{00}^{00}(u)+R_{10}^{10}(u))}=1,&\\&
\nn
\frac{R_{11}^{11}(u)([R_{11}^{11}(u)]^2-[R_{10}^{10}(u)]^2)-R_{01}^{01}(u)([R_{00}^{00}(u)]^2-[R_{01}^{01}(u)]^2)}{
(R_{00}^{00}(u)+R_{10}^{10}(u))(R_{01}^{10}(u)
R_{10}^{01}(u)+R_{00}^{11}(u)R_{11}^{00}(u))+(d_{-})R^{11}_{00}(u)R_{10}^{01}(u)(R_{11}^{11}(u)+R_{01}^{01}(u))}=1.&
\ena
We can see that two relations above can be obtained one from other by
the transformations  of the matrix elements -   $R_{ij}^{kr}\to
R_{\bar{i}\bar{j}}^{\bar{k}\bar{r}} $,  with $\bar{i}=(i+1)\;
\mathrm{mod}\; 2$. The constants $d_{\pm}$ also are connected one with other, as one can define (\cite{KhS}) \\
$d_{+}=\frac{3{R_{00}^{00}}'(0)-{R_{11}^{11}}'(0)-{R_{01}^{01}}'(0)-{R_{10}^{10}}'(0)-
{R_{01}^{10}}'(0)-{R_{10}^{01}}'(0)}{{R_{11}^{00}}'(0)}$,
$d_{-}=\frac{3{R_{11}^{11}}'(0)-{R_{00}^{00}}'(0)-{R_{01}^{01}}'(0)-{R_{10}^{10}}'(0)-
{R_{01}^{10}}'(0)-{R_{10}^{01}}'(0)}{{R_{00}^{11}}'(0)}$.

For the  cases of  next  $n$-s ($n\geq 4$)  the commutativity of the transfer matrices $\tau_n$ also are
ensured by obtained relations (\ref{t2h}, \ref{s12}, \ref{nsym-r}), so they entirely define the $R$-matrices for integrable models.
The 
obtained relations intend that for the general non-homogeneous case there are four independent functions (one of them always can be taken as unity) and three arbitrary
 constants by means of which all the matrix elements can be expressed, as it was stated already  by direct solving the YBE (see the works \cite{KhS,SHY} and the citations therein).
   As it is discussed therein, 
 the following YBE  
  (\ref{ybe1})
 $ \bar{R}(u,v)R(u)R(v)=R(v)R(u)\bar{R}(u,v)$, define uniquely the intertwiner matrix $\bar{R}(u,v)$,
   and there are no additional constraints on   $\bar{R}(u,v)$ and $R(u)$, and it appears  that $R(u)=\bar{R}(u,0)$.  The
  familiar elliptic, trigonometric and rational parameterizations can be obtained letting $\bar{R}(u,v)=\bar{R}(u-v)$.

 \end{small}}
{\begin{figure}[b]
\unitlength=12pt
{\includegraphics[width=\textwidth]{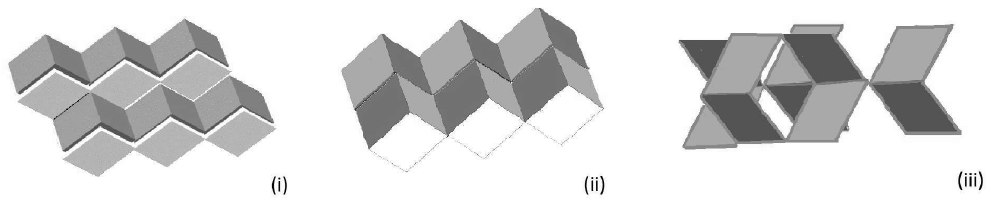}}\hfil
\vspace{-15cm}
\caption{{{“Connected”} (or {combined}) star-triangle relations: \it the corresponding transfer matrices and the plaquette
 weights;
 (i) product of two
 transfer matrices -  darker  stripes shifted by  a link  belong to two  vertically disposed fragments of 2d transfer matrices, (ii)
   checkerboard arrangement of plaquette weights in a two-dimensional  transfer matrix, (iii)  the weight plaquettes on a fragment of the product of two
 transfer matrices - at the  junctions between the plaquettes, the edges of  horizontal or vertical intertwiner weights can be located.}}\label{fig3}
\end{figure}}
%

\begin{small}

\end{small}

\end{document}